\documentclass[12pt]{article}

\usepackage{columbia}

\numberwithin{equation}{section}

\begin{document} 
\title{\textbf{Platonic solids back in the sky:\\
Icosahedral inflation}}
\author{Jonghee Kang}
\author{Alberto Nicolis}
\affil{Physics Department and Institute for Strings, Cosmology and Astroparticle Physics,\\ \vspace{.15cm}
Columbia University, New York, NY 10027, USA}
\date{}
\maketitle 

\begin{abstract}
We generalize the model of solid inflation to an anisotropic cosmic solid. Barring fine tunings, the observed isotropy of the cosmological background and of the scalar two-point function  isolate the icosahedral group as the only possible symmetry group of such a solid.
In such a case, higher-point correlation functions---starting with the three-point one---are naturally maximally anisotropic, which makes the standard detection strategies highly inefficient and calls for a dedicated analysis of CMB data. The tensor {\em two}-point function can also be highly anisotropic, but only in the presence of sizable higher-derivative couplings.
\end{abstract}

\section{Introduction}
The universe appears isotropic on large scales, and it is thus natural to assume that whatever it was that fueled primordial inflation, it was an isotropic system. It is interesting, however, to analyze critically this assumption. Observations tell us that the cosmological background and the spectrum of scalar perturbations are isotropic, but they do not tell us anything about higher-point  correlation functions or about tensor modes, for the simple reason that we have not detected them yet.

This raises the following question: can one have a physical system driving inflation whose dynamics are intrinsically anisotropic---perhaps maximally so---but that nevertheless yields an isotropic background and an isotropic scalar spectrum of perturbations? We are not interested in systems for which one can tune coefficients in the Lagrangian in order to achieve the desired degree of isotropy, but rather in systems whose symmetries are so powerful as to enforce such an isotropy, leaving open the possibility of anisotropic signals for other observables.

To make the discussion more concrete, let's consider the cubic group. This is the discrete subgroup of rotations that maps a cubic lattice into itself. Calling $\hat x$, $\hat y$, $\hat z$ the lattice's preferred directions, the cubic group is simply the set of permutations of $\hat x$, $\hat y$, $\hat z$ as well as single-axis inversions $\hat x \to - \hat x$, etc. Barring fine tunings, the dynamics of an homogenous system with this symmetry group---such as a cubic crystal in the continuum limit---in general will not be isotropic. However, certain observables are forced to be. In particular, because of cubic symmetry, any two-index tensor associated with the lattice must take the form
\be
T^{ij} \propto \hat x^i \hat x^j + \hat y^i \hat y^j + \hat z^i \hat z^j = \delta^{ij} \; .
\ee
On the other hand, with more indices there are structures that are invariant under the cubic group only, and can in principle lead to observable anisotropies. For instance, at the four-index level, the  tensor structure
\be
\hat x^i \hat x^j \hat x^k \hat x^l + \hat y^i \hat y^j \hat y^k \hat y^l+\hat z^i \hat z^j \hat z^k \hat z^l
\ee
is invariant under the cubic group, but cannot be rewritten in terms of Kronecker deltas only.

Solids are natural candidates for considering discrete subgroup of rotations, and for this reason we will elaborate on the above ideas in the context of  solid inflation \cite{Endlich:2012pz} , which we  review in the next section. In solid inflation, inflation is driven by a solid's stress-energy tensor $T^{\mu\nu}$. For the background evolution to be isotropic, one needs an isotropic $T^{ij}$ on the ground state of the solid. However, $T^{ij}$ in general is invariant only under the symmetries of the solid under consideration, which restricts the number of possible symmetry groups to those whose invariant two-index tensors are accidentally isotropic. As we saw above, the cubic group has this property.

Moving away from the background, the constraints become more severe.
The fluctuating degrees of freedom are the solid's phonons, which can be parametrized by a 3-vector field $\vec \pi(x)$, and the metric perturbations. To discuss possible anisotropies of scalar correlation functions, it is sufficient to focus on the phonons: the longitudinal one mixes with the scalar metric perturbations, and so any anisotropies in its dynamics will be reflected in scalar correlation functions. In particular, the two-point function is determined by the phonons' quadratic Lagrangian, which takes the general form
\be
\lagr_2 = A_{ij} \dot{\pi}^i \dot{\pi}^j  + B_{ijlm} \pa^i \pi^j \pa^l \pi^m \; , 
\ee
where $A^{ij}$ and $B^{ijlm}$ are tensors that are invariant under the symmetry group of the solid. We see that for the scalar two-point function to be isotropic, one also needs the invariant four-index tensors to be isotropic. As we saw above, the cubic group does not pass this test.

We will show in sect.~\ref{ico group} that the only discrete subgroup of rotations with the above properties is the icosahedral group---the symmetry group of an icosahedron. The natural question now is which observables are going to exhibit the anisotropies associated with such an icosahedron: there should be many preferred directions in the sky (20, 30, or 12, depending on whether one counts the faces, edges, or vertices), which should show up in correlation functions. The question can be approached once again in terms of invariant tensors. 
The scalar three-point function is determined by the cubic Lagrangian for the phonons, which now can involve a six-index tensor:
\be
\lagr_3 \supset T^{ijklmn} \pa_i \pi_j \pa_k \pi_l \pa_m \pi_n \; .
\ee
We will show that the icosahedral group allows for the anisotropic invariant tensor
\be
T^{ijklmn}_{\textrm{aniso}} \propto 2(\gamma + 2)  \, \delta^{ijklmn} + (\gamma + 1)  \l( \delta^{ijkl}\delta^{mn}\delta^{m\, i+1} + \cdots \r) + \l( \delta^{ijkl}\delta^{mn}\delta^{m\, i-1} + \cdots \r) \; , \nonumber
\ee
where $\gamma$ is the golden ratio, the six-index and four-index deltas are nonzero only if {\em all} their indices take the same value, and $i+1$ and $i-1$ are to be interpreted modulo 3, that is, $3+1 \to 1$ and $1-1 \to 3$. 
This makes the three-point function potentially highly anisotropic. In fact, we will show that there is a choice of Lagrangian coefficients for which the three-point function is {\em completely} anisotropic, in the sense that it has exactly vanishing overlap with any three-point function template associated with  isotropic models.

Anisotropies can show up in the tensor spectrum as well. The reason is that the quadratic Lagrangian for tensor modes also involves a six-index invariant tensor,
\be
\lagr_2^{(\gamma)} \supset C^{ijklmn} \, \pa_i \gamma_{jk} \, \pa_l \gamma_{mn} \; .
\ee
However, we will see that such six-index tensor can receive anisotropic contributions only from higher-derivative terms in the Lagrangian. It is consistent within our effective theory to assume that these are so large as to yield order-one anisotropies in the tensor spectrum, and in the examples we considered we found no indication that this disrupts the technical naturalness of the effective theory, but we have not investigated the question systematically. We leave this for future work. Notice that a strong anisotropy in the tensor spectrum can in principle reconcile tensions between a large tensor signal in a small patch of the sky and little or no signal in a whole-sky average, like the original---then evaporated---BICEP2/Planck tension.

Finally, we should emphasize that when we talk about `solids' we do not mean systems with an underlying crystal structure, but rather  continuous, homogeneous solid media, which can be more symmetric than crystals in the continuum limit. For instance, there is no crystal with icosahedral symmetry group, but it is perfectly consistent to assume such a symmetry for a continuous medium (in fact, there are {\em quasi}-crystals with icosahedral symmetry \cite{steinhardt}.) Even though the solids of everyday life are not homogeneous at microscopic scales, there is no a priori reason why there couldn't exist (perhaps strongly coupled) field theories that at finite density exhibit perfectly homogeneous solid-like states. If one is uncomfortable with such an assumption, one can regard our inflationary model simply as a system of three scalar fields with certain symmetries. As we now review, the low-energy effective field theory is the same, which makes the difference between the two viewpoints unsubstantial.

\section{Solid inflation}

From an effective field theory standpoint, the mechanical deformations of an homogeneous solid can be described in terms of three scalar fields $\phi^I(x)$ ($I=1,2,3$)  \cite{DGNR}, whose  expectation values in the ground state of the solid are 
\be \la{vev}
\langle \phi^I \rangle  =  x^I
\ee
and whose Lagrangian is invariant under the shift symmetries
\be
\phi^I \to \phi^I + a^I \, , \quad\quad a^I = \textrm{const} \la{shift}
\ee
(see also \cite{gruzinov, sigurdson} for alternative approaches.)
The $\phi^I$'s can be regarded as the comoving coordinates of the solid's volume elements. By Poincar\'e- and shift-invariance, to lowest order in derivatives  the Lagrangian must take the form
\be
{\cal L} = F\big( B^{IJ} \big) \; , \qquad B^{IJ} \equiv \partial_\mu \phi^I \partial^\mu \phi^J \; ,
\ee
where $F$ is an a-priori generic function, determined by the solid's equation of state.
For a solid with symmetry group $G \subset SO(3)$, one also demands that the Lagrangian be invariant under the internal rotations
\be
\phi^I \to {O^I}_J \phi^J \, , \quad\quad {O^I}_J \in G \la{rot}  \; ,
\ee
which restricts the form of $F$.

For instance, in the case in which $G$ is the full $SO(3)$---which is the case extensively studied in \cite{Endlich:2012pz}---$F$ can only depend on three invariants, e.g.
\be
[B] \; , \qquad [B^2] \;, \qquad [B^3] \; , \label{invariants}
\ee
where the square brackets denote the trace of the matrix within. Any other rotationally invariant function of $B^{IJ}$ can be expressed in terms of these, e.g.
\be
\det B = \sfrac16 \big([B]^3 - 3[B] [B^2] +2 [B^3]    \big) \; .
\ee

Upon minimally coupling the solid to gravity,
the form of $F$ is restricted further by demanding that the solid be able to drive near exponential inflation. In order for that to happen, one needs a solid that can be stretched by a large exponential factor without changing too much its physical properties, such as its energy density. Such a behavior is of course unlike that of any standard solid we know of, but it can be achieved by imposing an approximate {\em internal} scale invariance \cite{Endlich:2012pz}
\be
\phi^I \to \lambda \, \phi^I \; . \label{scale}
\ee 
Focusing again on the $SO(3)$ invariant case, to implement this symmetry it is useful to organize the three invariants \eqref{invariants} as
\be
X = [B] \; , \qquad Y= \frac{[B^2]}{[B]^2} \; , \qquad Z= \frac{[B^3]}{[B]^3} \; . \label{XYZ}
\ee
$X$ depends on the overall normalization of $B$, but $Y$ and $Z$ do not, and as a result $Y$ and $Z$ are invariant under the transformation  \eqref{scale}. The requirement of approximate scale invariance thus translates into a weak dependence of $F(X,Y,Z)$ on $X$. In particular, by evaluating the solid's stress-energy tensor on an FRW background, one finds \cite{Endlich:2012pz}
%
%
%
\be
\rho = -F, \qquad p = F-\tfrac{2}{a^2} F_X \qquad \qquad(X = {3}/{a^2}) \; ,
\ee
which yields the slow-roll parameter
\be
\epsilon \equiv -\fr{\dot{H}}{H^2}  =\fr{3}{a^2}\fr{F_X}{F} = \fr{\pa \textrm{log} F}{\pa \textrm{log} X} \; .
\ee
Of course ``slow-roll" here is a bad characterization, because nothing is rolling, slowly or otherwise: the background configurations for our $\phi^I$'s only depend on the {\em spatial} coordinates. But slow-roll parameters like $\epsilon$ and the higher order ones can also be defined geometrically, without any reference to rolling fields, purely in terms of the time-dependence of the Hubble scale $H$. We will adopt these geometric definitions---as we did above for $\epsilon$---and still use the standard slow-roll nomenclature.

In the presence of perturbations, the scalars and the metric are
\be
\phi^I = x^I + \pi^I(x) \; , \qquad g_{\mu\nu} = g_{\mu\nu}^{\rm FRW}(t) + \delta g_{\mu\nu}(x) \; ,
\ee
where $g_{\mu\nu}^{\rm FRW} = {\rm diag}(-1, a^2, a^2, a^2)$ is the standard FRW metric. At distances much shorter than the Hubble radius, one can neglect the metric perturbations and identify $\vec \pi(x) $ with the phonon field. Expanding the solid Lagrangian to quadratic order, one finds two parameters $c_L$ and $c_T$ characterizing the longitudinal and transverse phonon propagation speed. These are determined by certain derivatives of $F$, evaluated on the background configuration. In particular, one finds the universal exact relation \cite{Endlich:2012pz}
\be \label{cL cT}
c_T^2 = \sfrac34 \big(1 + c_L^2 - \sfrac23 \epsilon + \sfrac13 \eta \big) \; ,
\ee
where $\eta \equiv \dot \epsilon/ \epsilon H $ is the second slow-roll parameter.

To study cosmological perturbations and compute their correlation functions, it is convenient to decompose the metric in an ADM fashion,
\be
ds^2 = -N^2 dt^2 + h_{ij} \big( dx^i + N^i dt \big) \big( dx^j + N^j dt\big) \; ,
\ee
and choose spatially flat slice gauge (SFSG), 
\be
h_{ij} = a(t)^2\exp{(\gamma_{ij})} \; , \qquad \pa_i \gamma_{ij} = \gamma_{ii} = 0 \; .
\ee
%
%
%
The curvature perturbation $\zeta$ can be defined in a gauge-invariant fashion, and in the above gauge it is related to the phonon field $\vec \pi$ by
\be
\zeta = \sfrac{1}{3} \vec \nabla \cdot \vec{\pi} \; .
\ee
Then, using standard cosmological perturbation theory, one can compute the correlation functions for scalar and tensor modes. At the two-point function level, the relevant observables are the scalar tilt, the tensor tilt, and the tensor-to-scalar ratio:
\begin{align}
n_S -1  & \simeq 2 \, \epsilon c_{L}^2 -5s -\eta \\
n_T -1 &  \simeq 2 c_{L}^2 \epsilon \\
r  & \simeq 16 \,  \epsilon c_L^5 \; ,
\end{align}
where $s$ monitors the time-dependence of $c_L$, $s \equiv \dot c_L/c_L H$. Particularly unusual predictions are the positivity of the tensor tilt---which would usually require a violation of the null energy condition---and the strong suppression of the tensor-to-scalar ratio in the slow sound speed limit, a factor of $c_L^4$ stronger than for standard single-field cases.


Expanding further the solid Lagrangian to cubic order, one finds that at leading order in slow-roll the phonon self-interactions take the form
\be \la{originalcubic}
\mathcal{L}^{(3)} = \mpl^2  a(t)^3 H^2 \fr{F_Y}{F}\Big\{ \tfrac{7}{81} (\pa_i \pi^i)^3 - \tfrac{1}{9} \pa_i \pi^i \pa_j \pi^k \pa_k \pi^j -\tfrac{4}{9} \pa_i \pi^i \pa_j \pi^k \pa_j \pi^k + \tfrac{2}{3} \pa_j \pi^i \pa_j \pi^k \pa_k \pi^i \Big\} \; .
\ee
The gravitational corrections to this are suppressed both in the de-mixing regime, $k \gg aH\epsilon^{1/2}$, and in the strong mixing  one, $k \ll aH\epsilon^{1/2}$, and one can argue that the cubic Lagrangian above is all one needs to compute the three-point function of curvature perturbations \cite{Endlich:2012pz}.

To leading order is slow roll, the result is
\ba \la{3ptoriginal}
\big \langle \zeta(\vec{k}_1) \zeta(\vec{k}_2) \zeta(\vec{k}_3) \big\rangle & \simeq & (2\pi)^3 \delta^3(\vec{k}_1+\vec{k}_2+\vec{k}_3)\fr{3}{32} \fr{F_Y}{F} \fr{H^4}{ \mpl^4}\fr{1}{\epsilon^3c_L^{12}} \nonumber
\\
&& \times\, \fr{Q(\vec{k}_1,\vec{k}_2,\vec{k}_3)U(k_1,k_2,k_3)}{k_1^3 k_2^3 k_3^3}
\ea
where
\ba
Q(\vec{k}_1,\vec{k}_2,\vec{k}_3) &\equiv&\fr{7}{81} k_1 k_2 k_3 -\fr{5}{27}\l(k_1 \fr{(\vec{k}_2\cdot\vec{k}_3)^2}{k_2 k_3}+k_2 \fr{(\vec{k}_3\cdot\vec{k}_1)^2}{k_3 k_1}+k_3 \fr{(\vec{k}_1\cdot\vec{k}_2)^2}{k_1 k_2} \r) \nonumber
\\
&&+\fr{2}{3}\fr{(\vec{k}_1\cdot\vec{k}_2)(\vec{k}_2\cdot\vec{k}_3)(\vec{k}_3\cdot\vec{k}_1)}{k_1 k_2 k_3}
\ea
and
\ba
U(k_1,k_2,k_3) &=& \fr{2}{k_1 k_2 k_3 \l(k_1 + k_2 + k_3 \r)^3} \Big\{ 3 \big( k_1^6 + k_2^6 + k_3^6 \big) + 20 k_1^2 k_2^2 k_3^2 \nonumber
\\
&&+18\big(k_1^4 k_2 k_3 + k_1 k_2^4 k_3 + k_1k_2k_3^4 \big) +12\big( k_1^3 k_2^3 + k_2^3 k_3^3 + k_3^3 k_1^3 \big) \nonumber
\\
&& 9 \big( k_1^5 k_2 + 5 \, \textrm{perms}  \big) + 12\big( k_1^4 k_2^2 + 5\,\textrm{perms}\big) \nonumber
\\
&&+ 18 \big( k_1^3 k_2^3 k_3 + 5\, \textrm{perms} \big) \Big\} \; .
\ea
Assuming $F_Y \sim F$, this has a potentially huge $f_{\rm NL}$,
\be
f_{\textrm{NL}} = -\fr{19415}{13122}  \fr{F_Y}{F} \fr{1}{\epsilon c_L^2} \sim \fr{1}{\epsilon c_L^2}  \; ,
\ee
but its most peculiar feature is probably its `shape' \cite{Babich:2004gb}---in particular, its purely quadrupolar angular dependence in the squeezed limit 
$k_3 \ll k_{1,2}$:
\be
\langle \zeta \zeta \zeta \rangle \propto  \fr{(1-3 \cos^2{\theta})}{k_1^3 k_3^3} \; , 
\ee
where $\theta$ is the angle between $\vec{k}_1$ and $\vec{k}_3$.

\section{Hunting for the right  symmetry group}\label{ico group}

We now want to generalize all of the above to a more general solid, invariant only under a discrete subgroup of rotations, which nonetheless  features the desiderata identified in the Introduction: an isotropic background stress-tensor, and an isotropic quadratic Lagrangian for the phonons. As we saw, at the mathematical level these requirements are equivalent to demanding that all invariant two-index and four-index tensors be fully isotropic for the symmetry group in question.

There is an infinite number of discrete subgroups of $SO(3)$, divided into two main classes: the crystallographic point groups and  non-crystallographic ones. Let's start with the former class.
A crystallographic point group is the symmetry group of a crystal system that can fill all of space. This means that the group has to map all the lattice points into one another, which is a stronger requirement than being simply a subgroup of rotations. Since there is only a finite number of  crystal systems---triclinic, monoclinic, orthorhombic, tetragonal, trigonal, hexagonal, and cubic---there is only a finite number of crystallographic point groups.
Except for the hexagonal one, all crystal systems can be defined in terms of their three primitive lattice basis vector, let's call them $\vec{a}$, $\vec{b}$, and $\vec{c}$. Then, the two-index tensor
\be
T^{ij} = a^i a^j + b^i b^j + c^i c^j
\ee
is invariant under the corresponding crystallographic point group. However, this tensor is not invariant under general $SO(3)$ rotations unless $\vec{a}$, $\vec{b}$, and $\vec{c}$ all have the same length and are all orthogonal to one another. So, only the cubic crystal survives.  Still, as already pointed out in the Introduction, the cubic group fails our test at the four-index level, because the tensor structure
\be
a^i a^j a^l a^m + b^i b^j b^l b^m + c^i c^j c^l c^m
\ee
is invariant under the cubic group but not under general $SO(3)$ rotations. We thus reach the conclusion that no crystallographic point group can meet both of our criteria
\footnote{The only possible exception to this argument is the hexagonal crystal, some of whose links are not primitive lattice vectors but rather suitable linear combinations thereof.  Still, one can easily show that certain two-index tensors that are invariant under the hexagonal group are not $SO(3)$ invariant, e.g.
\be
T^{ij} \propto b_1^i b_1^j + b_2^i b_2^j + b_3^i b_3^j \; ,
\ee
where $b_1 = (1,0,0)$, $b_2 = (1/2,\sqrt{3}/2,0)$, $b_3 = (-1/2,\sqrt{3}/2,0)$.}.

The non-crystallographic point groups are the icosahedral group, the infinitely many $C_n$ groups ($n$-fold rotations about a given axis), and the extensions of $C_n$ that include some kind of reflection. In the two last cases, already at the two-index level we can easily construct invariant tensors that are not $SO(3)$ invariant: for instance, the projector onto the plane perpendicular to the rotation axis.

So, all our bets are on the icosahedral group---the symmetry group of the icosahedron. The icosahedron has 20 triangular faces, 30 edges, and 12 vertices, and there are 60 proper rotations that maps it into itself. 
\begin{figure}
\centering
\includegraphics[width=6cm]{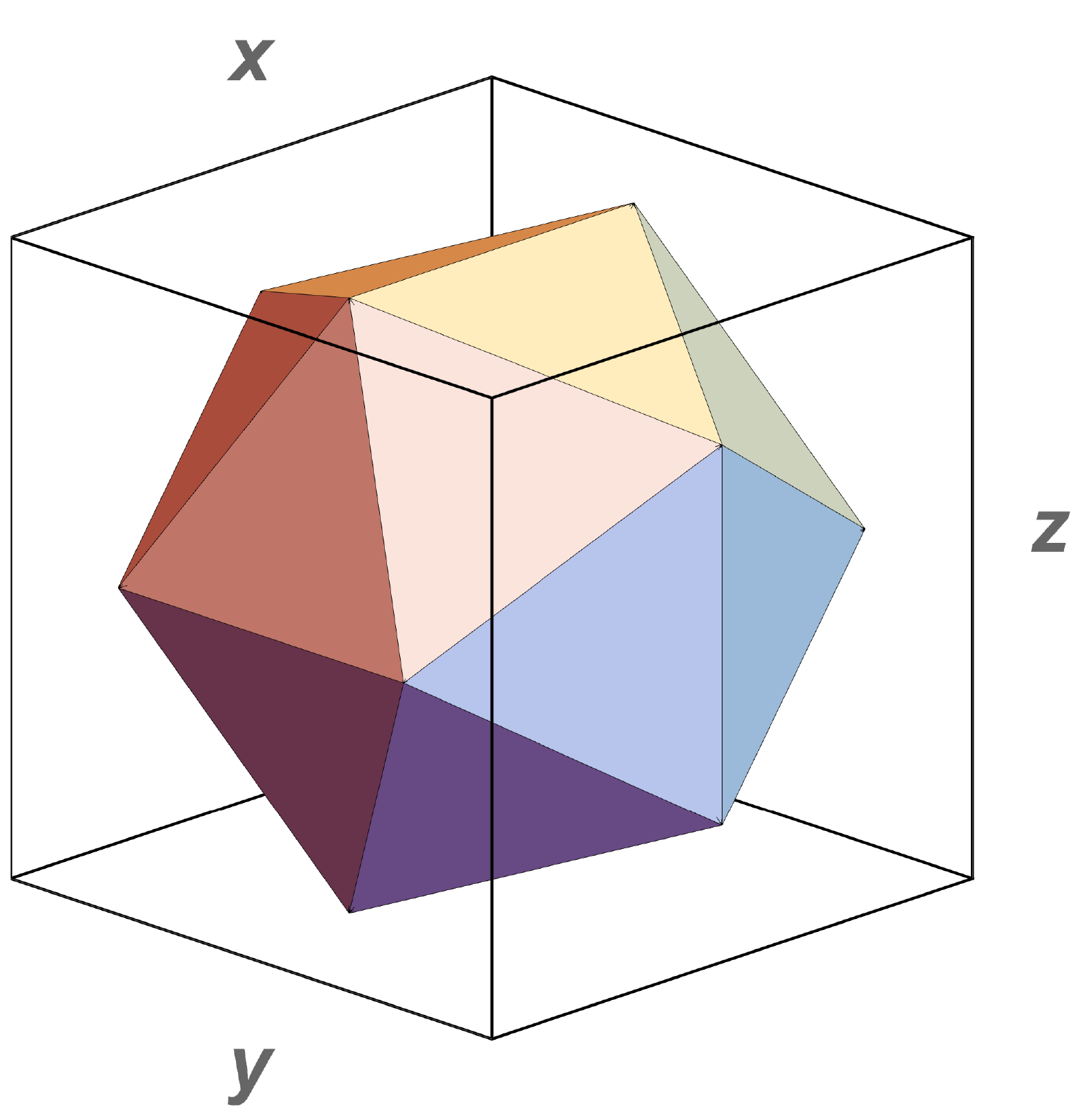}  
\caption{\em\small The relative orientation between our coordinate system and the icosahedron discussed in the text.\label{ico}}
\end{figure}
Following  \cite{litvin}, we orient our cartesian axes so that the icosahedron has two sides parallel to $x$, two parallel to $y$, and two parallel to $z$, as depicted in fig.~\ref{ico}. In this case, the coordinates of the vertices are (up to an overall rescaling)
\be
(\pm \gamma, \pm 1, 0) \; , \qquad (0, \pm \gamma, \pm 1) \; , \qquad  (\pm1, 0 , \pm \gamma)  \; ,
\ee
where $\gamma=\fr{\sqrt{5}+1}{2}$ is the golden ratio.

We can find the invariant tensors in the following way. By definition, the two-index invariant tensors should satisfy 
\be \la{2ind}
T^{i j} = T' {}^{ij} \equiv {I^{i}}_a\, {I^{j}}_b\,T^{ab} 
\ee
for each rotation matrix $I$ that belongs to the icosahedral group (we refer the reader to \cite{litvin} for the explicit form of the rotation matrices). Given that there are $60$ elements in the icosahedral group and $9 = 3\times3$  entries in $T^{ij}$, eq.~\eqref{2ind} can be interpreted as a system of $60\times9$ linear equations for the entries of $T^{ij}$. The solutions  are
\footnote{We used Mathematica to solve this linear system and those that follow.}
\be
T^{11} = T^{22} = T^{33}, \qquad \textrm{all other} \,\, T^{ij} = 0 \; .
\ee
That is, all two-index tensors that are invariant under the icosahedral group have the form
\be
T^{ij} \propto \delta^{ij} \; ,
\ee
and are therefore fully isotropic.

We can apply the same logic to the four-index invariant tensors:
\be \la{4ind}
T^{i j l m} = T' {}^{ijlm} \equiv  {I^{i}}_a\, {I^{j}}_b\, {I^l}_c\,{I^m}_d\,T^{abcd} \; .
\ee
In this case, we have $60\times81$ linear equations for the $81 = 3^4$ entries of $T^{ijlm}$, and the solutions are
\ba
&&T^{1122} = T^{1133} = T^{2211} = T^{2233} = T^{3311} = T^{3322} \nonumber
\\
&&T^{1212} = T^{1313} = T^{2121} = T^{2323} = T^{3131} = T^{3232} \nonumber
\\
&& T^{1221} = T^{1331} = T^{2112} = T^{2332} = T^{3113} = T^{3223} \nonumber
\\
&& T^{1111} = T^{2222}  = T^{3333} = T^{1122} + T^{1212} + T^{1221} \nonumber
\\
&& \textrm{all other} \,\, T^{ijlm} = 0 \; .
\ea
These conditions can be rewritten compactly using only Kronecker deltas, which shows that all four-index invariant tensors are fully isotropic as well:
\be
T^{ijlm} = A \, \delta^{ij}\delta^{lm} + B \, \delta^{il}\delta^{jm} + C \, \delta^{im} \delta^{jl}
\ee
for arbitrary $A$, $B$, and $C$.

Finally, let us consider the six-index invariant tensors. Following the same logic, we now have $60\times3^6 = 60\times 729$ equations
\be \la{6ind}
T^{i j k l m n} = T' {}^{ijklmn} \equiv {I^{i}}_a\, {I^{j}}_b\, {I^k}_c\,{I^l}_d\,{I^m}_e\,{I^n}_f\,T^{abcdef} \; .
\ee
On top of the isotropic solutions, schematically of the form $T \sim \delta \delta \delta$, we find an anisotropic one:
\be
T^{ijklmn}_{\textrm{aniso}}  = 2(\gamma + 2) \, \delta^{ijklmn} + (\gamma + 1) \l( \delta^{ijkl}\delta^{mn}\delta^{m\, i+1} + \cdots \r) + \l( \delta^{ijkl}\delta^{mn}\delta^{m\, i-1} + \cdots \r)
\ee
where the dots stand for all other combinations of four  and two indices out of six, the delta tensors with more than two indices are 1 only if {\em all}  those indices take the same value, and $i+1$ and $i-1$ are to be interpreted modulo 3, that is $ 3+1 = 1$ and $ 1-1 = 3$. 

It is worth mentioning that one could have derived the invariant tensors above in a perhaps more intuitive fashion, by using as building blocks the 12 vectors $\vec v_a$  ($a=1, \dots, 12$) that define the icosahedron's vertices. Clearly, by taking suitable tensor products and summing over all  the vertices, e.g.
\be
T^{i_1 \cdots i_n} = \sum_a v_a^{i_1} \cdots v_a^{i_n}  \; , 
\ee
one gets tensors that are invariant under the icosahedral group. However, it is not obvious that one can get {\em all} the invariant tensors in this way. Our brute-force analysis above settles the question (and the answer is `yes', at least up to the six-index level, if one includes tensor products of lower-order tensors as well.)


In conclusion, the icosahedral group has exactly the properties that we are after: all its two-index and four-index invariant tensors are isotropic, whereas its six-index ones are not. And it is the only subgroup of $SO(3)$ with these properties. 
From now on we will thus focus on a variant of solid inflation with icosahedral symmetry, which we dub `icosahedral inflation'. 
As already emphasized, the anisotropy of the six-index invariant tensors translates into an anisotropy of the scalar three-point function. Our goal now is to compute such a three-point function.

\section{Cubic Lagrangian of scalar modes} \label{cubiclag}
An anisotropic invariant six-index tensor can induce anisotropies in the scalar three-point function through the trilinear phonon interaction 
\be
\lagr_3 \propto T^{ijklmn}_{\textrm{aniso}} \pa_i \pi_j \pa_k \pi_l \pa_m \pi_n \; .
\ee
However, to figure out the most general structure of the cubic Lagrangian compatible with our symmetries requires some work. In the $SO(3)$-invariant version of solid inflation, this task was straightforward: the full solid's Lagrangian only depends on the three invariants \eqref{invariants}, each of which can be expanded in perturbations about the background solution, up to any desired order. In icosahedral inflation, we face the problem of classifying the allowed invariants of $B^{IJ}$. Since $B^{IJ}$ starts at zeroth order in $\vec \pi$,
\be
B \sim 1 + \pa \pi + \pa \pi \pa \pi \; ,
\ee
to expand the Lagrangian to any given order in $\vec \pi$---cubic, in our case---we need to consider all orders in $B^{IJ}$.
However, at high orders, in principle we have to include more and more invariants,
\be
T_{I_1 J_1 \cdots   I_n J_n} \, B^{I_1 J_1} \cdots B^{I_n J_n} \; ,
\ee
where $T$ is a generic $2n$-index tensor with icosahedral symmetry. We are not aware of any simplifying property of the icosahedral group analogous to the $SO(3)$ statement that an arbitrary invariant of $B^{IJ}$ can be written as a non-linear function of the three fundamental invariants \eqref{invariants}. Clearly, the number of independent invariants cannot be more than the number of independent components of $B^{IJ}$---six---but using the individual components of $B^{IJ}$ would make our computations messy and unreadable.

To get around this problem, we can work directly with the {\em fluctuation} of $B^{IJ}$ about its background, but, as we will see below, we will have to be careful about the non-linearly realized symmetries
\footnote{In the standard effective field theory of inflation \cite{Cheung:2007st}, it is straightforward to write the action directly in terms of the metric perturbations in unitary gauge, by using for instance~$\delta g^{00} \equiv g^{00}+1$. As emphasized in \cite{Endlich:2012pz}, in solid inflation the analogous variable in unitary gauge would be $\delta g^{ij} = g^{ij}-\delta^{ij}/a^2$, but this, unlike the full $g^{ij}$, does not transform nicely under the residual time diffeomorphisms, because the background $\delta^{ij}/a^2$ does not.}.
In SFSG gauge, our building block up to cubic order is \ba
B^{ij} &=& -\fr{1}{N^2}\l( \dot{\pi}^i - N^k \pa_k \phi^i \r) \l( \dot{\pi}^j - N^k \pa_k \phi^j \r) + h^{km} \pa_k \phi^i \pa_m \phi^j \nonumber
\\
&\simeq& \fr{\delta^{ij}}{a^2} + \fr{1}{a^2} \l( \pa^i \pi^j + \pa^j \pi^i \r) + \fr{1}{a^2} \pa_k \pi^i \pa^k \pi^j - \l( \dot{\pi}^i - N^i \r) \l( \dot{\pi}^j - N^j \r) \nonumber
\\
&& +  \l( \dot{\pi}^i - N^i \r) N^k \pa_k \pi^j +  \l( \dot{\pi}^j - N^j \r) N^k \pa_k \pi^i + 2 \delta N  \l( \dot{\pi}^i - N^i \r) \l( \dot{\pi}^j - N^j \r) + \cdots \nonumber
\\
& \equiv & \fr{1}{a^2} \big(\delta^{ij}+ \pi^{ij}\big)
\ea
where we defined $\pi^{ij}$ as the fluctuating part of $B^{ij}$, and we stopped differentiating between the internal $I,J,\ldots$ indices and the spacial $i,j,\ldots$ ones (the reason is that  the background $\langle \phi^I \rangle= x^I$ breaks spatial rotations and internal ones down to the diagonal combination.) 
Were we to expand the full non-linear action up to cubic order, we would have
\be
F(B^{ij}) = F_0 + \fr{\pa F}{\pa B^{ij}} \Big|_0\,  \fr{\pi^{ij}}{a^2}  +\fr{1}{2!} \fr{\pa^2 F}{\pa B^{ij}B^{kl}} \Big|_0  \, \fr{\pi^{ij} \pi^{kl}}{a^4} +\fr{1}{3!} \fr{\pa^3 F}{\pa B^{ij} B^{kl}B^{mn}} \Big|_0  \fr{\pi^{ij} \pi^{kl} \pi^{mn} }{a^6}+ \cdots \; , 
\ee
where the subscript zeros mean `evaluated on the background'. By the background Friedmann equations, the first derivative of $F$ with respect to $B^{ij}$ can be related to $\epsilon$ \cite{Endlich:2012pz}:
\be
\fr{\pa F}{\pa B^{ij}} \Big|_0 = \epsilon \, \sfrac13 a^2 F_0 \, \delta_{ij} \; .
\ee
This result was derived for the original solid inflation model assuming $SO(3)$ invariance, but in the Appendix we prove that it holds for our icosahedral inflation case as well.
The higher derivatives of $F$ do not enter the Friedmann equations, and therefore cannot be related simply to other background quantities. As we will see below, they do obey constraints coming from the non-linearly realized symmetries, but for the moment we can just parametrize them as the most general icosahedral-invariant tensors with the right index-permutation symmetries ($i \leftrightarrow j$, $(ij) \leftrightarrow (kl)$, and so on). Since the factors of $\pi^{ij}$ they are contracted with have precisely the same permutation symmetries, we can simply write
\begin{align}
F(B^{ij}) = & \,  F_0 \cdot \big[ 1 + \sfrac13 \epsilon \,  \pi^{ii} + \alpha_4 \big( \delta_{ij} \delta_{kl} + \beta_1 \delta_{ik}\delta_{jl} \big) \pi^{ij}\pi^{kl}  \label{expanding F} \\
& + \alpha_6 \big( \delta_{ij}\delta_{kl}\delta_{mn}  + \beta_2 \delta_{ij}\delta_{km}\delta_{ln}+ \beta_3 \delta_{ik}\delta_{jn}\delta_{lm} + \beta_4 \, T^{ijklmn}_{\textrm{aniso}} \big)\pi^{ij}\pi^{kl} \pi^{mn} + \cdots \big] \; , \nonumber
\end{align}
where the $\alpha$'s and $\beta$'s are generic dimensionless coefficients, with a weak time-dependence that can be neglected to lowest order in slow-roll.

If for the moment we ignore the metric perturbations $\delta N$ and $N^i$, then $\pi^{ij}$ is simply
\be
\pi^{ij} = \l( \pa^i \pi^j + \pa^j \pi^i \r) +  \pa_k \pi^i \pa_k \pi^j - a^2 \, \dot{\pi}^i \dot{\pi}^j \; ,
\ee 
and isolating  the different orders in $\vec \pi$  in the action above is immediate.
At the quadratic level we get 
\ba
{\cal L}_2 &=& F_0 \cdot \big[ -\sfrac13 \epsilon a^2 \, \dot{\vec \pi}\, {}^2+ \l( \sfrac13 \epsilon  + {2 \alpha_4 \beta_1} \r) \l( \pa_i \pi_j \r)^2 +  2 \alpha_4 ( 2 + {\beta_1} ) \, ( \pa_i \pi^i )^2 \, \big] \nonumber
\\
&\equiv& -\sfrac13 \epsilon a^2 F_0 \cdot \l [ \dot{\vec \pi}\, {}^2 - {c_T^2} \, \fr{(\pa_i \pi_j)^2}{a^2} - {(c_L^2 - c_T^2)}\,  \fr{ (\pa_i \pi^i)^2}{a^2} \, \r] \; ,
\ea
where
\ba
c_T^2 &=& 1 + \fr{6\alpha_4 \beta_1}{\epsilon}
\\
c_L^2 &=& 1+ \fr{12\alpha_4 ( 1+ \beta_1 )}{\epsilon}
\ea
are the transverse and longitudinal phonon speeds. 
For these to be  between $0$ and $1$, we need both  $\alpha_4$ and $\alpha_4 \beta_1$ to be small,
\be \label{alphabeta}
\alpha_4  ,  \alpha_4 \beta_1 = {\cal O}(\epsilon) \; ,
\ee
in analogy with the  $F_Y + F_Z = {\cal O}(\epsilon) \cdot F$ requirement of the original solid inflation case \cite{Endlich:2012pz}. In fact, in the Appendix  we prove that the two propagation speed are related by the same constraint as in solid inflation, eq.~\eqref{cL cT}, so that---as anticipated---up to quadratic order in  perturbations our model is indistinguishable from solid inflation.

Eq.~\eqref{alphabeta} implies that, to lowest order in slow-roll, only the second line in \eqref{expanding F} contributes to the cubic Lagrangian:
\ba \la{cubiclag1}
\lagr_3 &\simeq& \alpha_6 F_0 \cdot  \l[  \l(8-\beta_3\r) \l( \pa_i \pi^i \r) ^3 + {4\beta_2} \l( \pa_i \pi^i \r) \l( \pa_j \pi^k \r)^2 +  \l( 4\beta_2 + 3\beta_3 \r) \pa_i \pi^i \pa_j \pi^k \pa_k \pi^j  \r. \nonumber
\\
&&\l. + \,{6\beta_3} \pa_j \pi^i \pa_j \pi^k \pa_k \pi^i +  8 \beta_4 \, T^{ijklmn}_{\textrm{aniso}} \pa_i \pi^j \pa_k \pi^l \pa_m \pi^n\r] \; .
\ea
But we are not done yet. Ref.~\cite{Endlich:2013jia} argued that the approximate internal scale invariance \eq{scale} manifests itself on the structure of the phonon self-interactions in the following way: the cubic action expanded about a phonon background $\vec \pi_0$ cannot correct  the quadratic action for the fluctuations if the background is isotropic, $\pa_i \pi_j^0 \propto \delta_{ij}$. Applying this requirement to our cubic action yields two constraints on the $\beta$'s,
\ba
72 + 28 \beta_2 + 12 \beta_3 + 48 (\gamma+2) \beta_4 &=& 0
\\
12 \beta_2 + 12 \beta_3 + 24 (\gamma + 2) \beta_4 &=& 0 \; ,
\ea
which allow us to eliminate $\beta_2$ and $\beta_4$,
\ba
\beta_2 &=& 3 \beta_3 - 18
\\
\beta_4 &=& \fr{9 - 2\beta_3}{\gamma+2} \; .
\ea

We are thus left with only two free coefficients, $\alpha_6$ and $\beta_3$, which from now on we will simply call $\alpha$ and $\beta$. In conclusion, the cubic Lagrangian for icosahedral inflation reads
\ba \la{cubiclag2}
\lagr_3 &=& {\alpha} F_0 \cdot \Big[ \l( 8 -\beta \r) \l( \pa_i \pi^i \r)^3 + \l(12 \beta - 72 \r) \pa_i \pi^i \l( \pa_j \pi^k \r)^2 +\l(15\beta-72 \r) \pa_i \pi^i \pa_j \pi^k \pa_k \pi^j   \nonumber
\\
&&  + 6\beta \, \pa_j \pi^i \pa_j \pi^k \pa_k \pi^i + \fr{8(9-2\beta)}{\gamma+2} \, T^{ijklmn}_{\textrm{aniso}}\, \pa_i\pi_j\pa_k\pi_l\pa_m\pi_n \Big]
\ea

Recall that in the original solid inflation model there was only one free coefficient at this order, $F_Y$, appearing as an overall factor in front the cubic Lagrangian \eq{originalcubic}---all the relative coefficients of the different terms were completely fixed. If we set our anisotropic structure to zero by setting $\beta = 9/2$, we recover precisely those ratios, and we get
\begin{equation}
{\alpha} = -\fr{2}{243}\frac{F_Y}{F} \qquad \qquad(\beta=9/2) \; .
\end{equation}
On the other hand, we will show in the next section that the choice $\beta = 8$ characterizes the {\em completely anisotropic} case, in the sense that the resulting three-point function has exactly zero overlap with all those that one could get from isotropic models. The cubic Lagrangian in this case is
\ba 
\lagr_3 &=& 8 {\alpha} F_0 \cdot \bigg[ 3 \, \pa_i \pi^i \l( \pa_j \pi^k \r)^2 +6 \,  \pa_i \pi^i \pa_j \pi^k \pa_k \pi^j   \nonumber
\\
&&  + 6 \,  \pa_j \pi^i \pa_j \pi^k \pa_k \pi^i -\fr{7}{\gamma+2} \, T^{ijklmn}_{\textrm{aniso}}\, \pa_i\pi_j\pa_k\pi_l\pa_m\pi_n \bigg]
\qquad\qquad (\beta=8) \; .
\ea

\section{The size and shape of non-gaussianities}

Like in the original case of solid inflation, and for the same reasons spelled out there \cite{Endlich:2012pz}, the leading trilinear interactions we need to consider to compute the scalar three-point function are the phonon self-interactions we wrote down above. That is, we can neglect interactions involving the metric perturbations. The computation of the three-point function parallels that in \cite{Endlich:2012pz}, with obvious modifications due the new tensor structures we have in the cubic Lagrangian. Neglecting the weak time-dependence of the scalar modes outside the horizon, the result is
%
%
%
\ba
\langle \zeta_1 \zeta_2 \zeta_3 \rangle &\simeq& (2\pi)^3 \delta^3 (\vec{k}_1 + \vec{k}_2 + \vec{k}_3) \times \nonumber
\\
&& ( -) \fr{9}{32} \fr{H^4}{\mpl^4} \cdot \fr{\alpha}{\epsilon^3 c_L^{12}}  \cdot \fr{Q(\vec{k}_1,\vec{k}_2,\vec{k}_3) U(k_1,k_2,k_3)}{k_1^3 k_2^3 k_3^3} \; ,
\ea
where
\ba
Q(\vec{k}_1,\vec{k}_2,\vec{k}_3)  &=& (8-\beta)\,k_1 k_2 k_3 + 6\beta \fr{\big(\vec{k}_1\cdot\vec{k}_2\big)\big(\vec{k}_2\cdot\vec{k}_3\big)\big(\vec{k}_3\cdot\vec{k}_1\big)}{k_1 k_2 k_3}  \nonumber
\\
&&+(9\beta-48) \, \Big( k_1 \fr{\big(\vec{k}_2\cdot\vec{k}_3\big)^2}{k_2 k_3}+k_2 \fr{\big(\vec{k}_3\cdot\vec{k}_1\big)^2}{k_3 k_1}+k_3 \fr{\big(\vec{k}_1\cdot\vec{k}_2\big)^2}{k_1 k_2} \Big)\nonumber
\\
&&+\fr{-16\beta+72}{\gamma+2} \fr{1}{k_1 k_2 k_3}\Big( 2(\gamma+2)k_1^i k_1^i k_2^i k_2^i k_3^i k_3^i +(\gamma+1)\big(k_1^i k_1^i k_2^i k_2^i k_3^{i+1} k_3^{i+1} \nonumber
\\
&&+k_1^i k_1^i k_2^{i+1} k_2^{i+1} k_3^{i} k_3^{i}+k_1^{i+1} k_1^{i+1} k_2^i k_2^i k_3^{i} k_3^{i}+4k_1^i k_1^i k_2^i k_2^{i+1} k_3^{i} k_3^{i+1} \nonumber
\\
&&+4k_1^i k_1^{i+1} k_2^i k_2^i k_3^{i} k_3^{i+1}+4k_1^i k_1^{i+1} k_2^i k_2^{i+1} k_3^{i} k_3^{i}\big)+\big(k_1^i k_1^i k_2^i k_2^i k_3^{i-1} k_3^{i-1}  \nonumber
\\
&&+k_1^i k_1^i k_2^{i-1} k_2^{i-1} k_3^{i} k_3^{i}+k_1^{i-1} k_1^{i-1} k_2^i k_2^i k_3^{i} k_3^{i}+4k_1^i k_1^i k_2^i k_2^{i-1} k_3^{i} k_3^{i- 1} \nonumber
\\
&&+4k_1^i k_1^{i-1} k_2^i k_2^i k_3^{i} k_3^{i-1}+4k_1^i k_1^{i-1} k_2^i k_2^{i-1} k_3^{i} k_3^{i}\big) \Big)
\ea
and
\ba
\fn{U}{k_1,k_2,k_3} &=& \fr{2}{k_1 k_2 k_3 \l(k_1 + k_2 + k_3 \r)^3} \Big\{ 3 \big( k_1^6 + k_2^6 + k_3^6 \big) + 20 k_1^2 k_2^2 k_3^2 \nonumber
\\
&&+18\big(k_1^4 k_2 k_3 + k_1 k_2^4 k_3 + k_1k_2k_3^4 \big) +12\big( k_1^3 k_2^3 + k_2^3 k_3^3 + k_3^3 k_1^3 \big) \nonumber
\\
&& 9 \big( k_1^5 k_2 + 5 \, \textrm{perms}  \big) + 12\big( k_1^4 k_2^2 + 5\,\textrm{perms}\big) \nonumber
\\
&&+ 18 \big( k_1^3 k_2^3 k_3 + 5\, \textrm{perms} \big) \Big\} \; .
\ea

The overall delta function leaves us  with only two independent momenta, say $\vec k_2$ and $\vec k_3$. Usually, because of isotropy, the absolute orientation of these two vectors does not matter, and one needs only three independent quantities to characterize the kinematical configuration: the magnitudes $k_2$ and $k_3$, and the relative angle $\theta$. For us, because of our anisotropies, the absolute orientation matters, and so we have to keep all the six components of $\vec{k}_2$ and $\vec{k}_3$. This complicates the analysis considerably. In particular, we cannot use the standard techniques of \cite{Babich:2004gb}.

A convenient parametrization of $\vec k_2$ and $\vec k_3$ is the following one. Define $\theta_2$ and $\phi_2$ as the standard polar and azimuthal angles of $\vec k_2$, but define $\theta_3$ and $\phi_3$ as the polar and azimuthal angles of $\vec k_3$ with respect to a primed coordinate system in which the $z'$ axis is along $\vec k_2$, and the $x'$ axis lies in the plane defined by the $z$ and $z'$ axes (see fig.~\ref{coordinates}).
The cartesian components of  $\vec{k}_2$ and $\vec{k}_3$ thus are
\ba
\vec{k}_2 &=& k_2 \, ( \sin{\theta_2}\cos{\phi_2},\,\sin{\theta_2}\sin{\phi_2},\,\cos{\theta_2} )
\\
\vec{k}_3 &=&k_3 \, ( \sin{\theta_2} \cos{\phi_2} \cos{\theta_3}+ \cos{\theta_2}\cos{\phi_2}\sin{\theta_3}\cos{\phi_3}-\sin{\phi_2}\sin{\theta_3}\sin{\phi_3},  \nonumber
\\
&&\sin{\theta_2} \sin{\phi_2} \cos{\theta_3}+ \cos{\theta_2}\sin{\phi_2}\sin{\theta_3}\cos{\phi_3}+\cos{\phi_2}\sin{\theta_3}\sin{\phi_3}, \nonumber
\\
&& \cos{\theta_2}\cos{\theta_3}-\sin{\theta_2}\sin{\theta_3}\cos{\phi_3} ) \; .
\ea
The advantage of this parametrization is that $\theta_3$ is  the relative angle between $\vec{k}_2$ and $\vec{k}_3$, and so any dependence on $\theta_3$ is perfectly consistent with isotropy. Anisotropies show up as a non-trivial dependence on $\phi_2$, $\phi_3$, and $\theta_2$.

\tdplotsetmaincoords{60}{105}
\pgfmathsetmacro{\rvec}{0.7}
\pgfmathsetmacro{\thetavec}{25}
\pgfmathsetmacro{\phivec}{45}
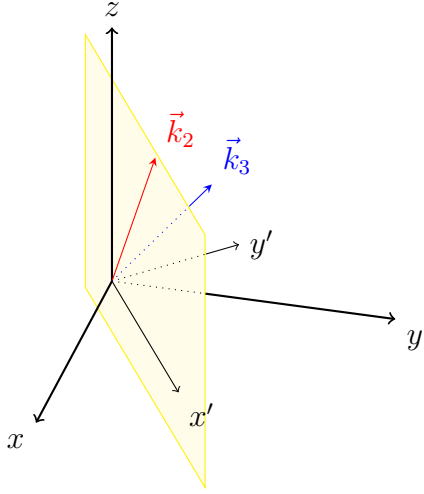
\begin{figure}[ht]
\centering
\begin{tikzpicture}[scale=3.9, tdplot_main_coords]

\coordinate (O) at (0,0,0);
\tdplotsetthetaplanecoords{\phivec}
\tdplotsetrotatedcoords{\phivec}{\thetavec}{0}
\tdplotsetrotatedcoordsorigin{(O)}
\tdplotsetcoord{P}{\rvec}{\thetavec}{\phivec}

\filldraw[ draw=yellow, fill=yellow!10] 
({0.7*cos(25) * cos(45)},{0.7*cos(25)*sin(45)},{-0.2 - 0.7*sin(25)}) 
-- ({0.7*cos(25) * cos(45)},{0.7*cos(25)*sin(45)},0.5)
-- (-{0.2*cos(25) * cos(45)},-{0.2*cos(25)*sin(45)},{0.2*sin(25) + 0.5 + 0.7*sin(25)}) 
-- ({-0.2*cos(25) * cos(45)},{-0.2*cos(25)*sin(45)},{-0.2 + 0.2*sin(25)})
-- cycle ; 
\draw[thick,->] (0,0,0) -- (1,0,0) node[anchor=north east]{$x$};
\draw[dotted] (0,0,0) -- (0,0.33,0);
\draw[thick,->] (0,0.33,0) -- (0,1,0) node[anchor=north west]{$y$};
\draw[thick,->] (0,0,0) -- (0,0,1) node[anchor=south]{$z$};

\draw[-stealth,color=red] (O) -- (P)node[anchor=south west]{$\vec{k}_2$};

\draw[tdplot_rotated_coords,->] (0,0,0) -- (.5,0,0) node[anchor=north west]{$x'$};
\draw[dotted,tdplot_rotated_coords] (0,0,0) -- (0,.37,0);
\draw[tdplot_rotated_coords,->] (0,.37,0) -- (0,.5,0) node[anchor=west]{$y'$};

\draw[dotted,color=blue,tdplot_rotated_coords] (0,0,0) -- ({.2*0.58},{.2*0.58},{.9*0.58});
\draw[-stealth,color=blue,tdplot_rotated_coords] ({.2*0.58},{.2*0.58},{.9*0.58}) -- ({0.2 * 0.75},{0.2*0.75},{0.9*0.75}) node[anchor=south west]{$\vec{k}_3$};
\end{tikzpicture}
\caption{\small\em The coordinate system defined in the text.  The $\hat{z}$, $\hat z'  =\hat k_2$, and $\hat x '$ axes all lie in the same plane.\label{coordinates}}
\end{figure}

Following the standard conventions for correlation functions of the Newtonian potential $\Phi$,
\ba
\Phi &=& \sfrac{3}{5}\zeta
\\
\big\langle\Phi(\vec{k_1})\Phi(\vec{k_2}) \big\rangle &=& (2\pi)^3 \delta^3 (\vec{k}_1 + \vec{k}_2) \fr{\Delta_\Phi}{k_1^3}
\\
\big\langle\Phi(\vec{k_1})\Phi(\vec{k_2})\Phi(\vec{k_3}) \big\rangle &=&  (2\pi)^3 \delta^3 (\vec{k}_1 + \vec{k}_2+ \vec{k}_2) f(\vec{k}_1,\vec{k}_2,\vec{k}_3) \la{modifiedconvention} \; ,
\ea
we define $f_{\rm NL}$ using equilateral configurations:
\be
f(\vec k_1, \vec k_2, \vec k_3) \big|_{\rm equil} = f_{\textrm{NL}} \fr{6 \Delta_\Phi^2}{k_1^6} \; .
\ee 
However,  the equilateral-triangle condition only fixes the relative angle $\theta_3$, and so in our case the resulting $f_{\rm NL}$ depends non-trivially on the other angles, $\phi_2$, $\phi_3$, and $\theta_2$.
To get a readable expression, we average $f_{\rm NL}$ over $\phi_2$ and $\phi_3$,
\be
\bar f_{\rm NL}(\theta_2) \equiv \fr{1}{4\pi^2} \int  d \phi_2 d \phi_3 \, f_{\textrm{NL}}(\theta_2, \phi_2, \phi_3) \; .
\ee
The remaining dependence on $\theta_2$ will still be a measure of anisotropy. For our three-point function, we get:
\ba
\Delta_\Phi &=& \fr{9}{100} \fr{H^2}{\mpl^2}\cdot \fr{1}{\epsilon c_L^5}
\\
f(\vec{k}_1,\vec{k}_2,\vec{k}_3)  &=& -\fr{15}{2} \fr{\alpha}{\epsilon c_L^2} \cdot \Delta_\Phi ^2 \cdot \fr{Q(\vec{k}_1,\vec{k}_2,\vec{k}_3) U(k_1,k_2,k_3) }{k_1^3 k_2^3 k_3^3}
\\
\bar f_{\rm NL}(\theta_2) &=& -   \fr{\alpha}{\epsilon c_L^2} \big[\sfrac{19415}{378}(\beta-8) + \sfrac{104135}{6048}(2\beta-9)P_6(\cos \theta_2) \big]  \; ,
\ea
where $P_6$ is the sixth order Legendre polynomial.

The typical size of $f_{\rm NL}$ is the same as in the standard solid inflation case, parametrically as big as $1/\epsilon c_L^2$ if one assumes $\alpha \sim 1$ (analogous to $F_Y \sim F$ for solid inflation). But clearly the most interesting feature here is the angular dependence of $f_{\rm NL}$: the appearance of $P_6$ in $\bar f_{\rm NL}(\theta_2)$ is the first indication that the case with $\beta = 8$ is a very special one, with a completely anisotropic $f_{\rm NL}$: if we average $\bar f_{\rm NL}$ over $\cos \theta_2$, which is equivalent to averaging the full $f_{\rm NL}$ over all angular variables,  we get zero.

We can go further and, following  \cite{Babich:2004gb}, consider the overlap between our three-point function and other `shapes'. This is defined as
\be \label{overlap}
\cos{}(f, f') \equiv \fr{f \cdot f'}{\sqrt{f \cdot f}\sqrt{f' \cdot f'}}
\ee
where
\be
f \cdot f' \equiv \sum\limits_{\vec{k}_i} f \big( \vec{k}_1,\vec{k}_2,\vec{k}_3 \big) f' \big( \vec{k}_1,\vec{k}_2,\vec{k}_3 \big) /  \big( \sigma_{k_1}^2\sigma_{k_2}^2\sigma_{k_3}^2\big) \; .
\ee
The sum runs over all triangles in momentum space, and is in fact an integral since the momenta are continuous variables. 

If for $f$ we take our shape (ignoring overall constant factors, which do not contribute to the overlap \eqref{overlap}),
\be
f ( \vec{k}_1,\vec{k}_2,\vec{k}_3 ) \to \fr{Q(\vec{k}_1,\vec{k}_2,\vec{k}_3) U(k_1,k_2,k_3)}{k_1^3 k_2^3 k_3^3} \; ,
\ee
and for $f'$ that coming from a general isotropic model,
\be
f' \to f' (k_1, k_2, k_3) \; ,
\ee
we find exactly vanishing overlap if $\beta=8$. Again, the reason is manifest if, when computing the angular integrals for the overlap \eqref{overlap}, we perform the integrals over  $\phi_2$ and $\phi_3$ first:
\begin{align}
\int d \! \cos \theta_2 \, d \phi_2 \, d \! \cos \theta_3  \, d \phi_3 \, f f'  = & \int d \! \cos \theta_3 \frac{32 \pi^2}{7} \frac{U\big(k_1,k_2,k_3 \big)}{k_1^4 k_2^4 k_3^4} k_2^2 k_3^2 \, f'(k_1,k_2,k_3) \\
& \times \int d \! \cos \theta_2 \big( (\beta-8) G_1(k_2,k_3,\theta_2,\theta_3 ) + (2\beta-9) G_2(k_2,k_3,\theta_2,\theta_3 )\big) \; , \nonumber
\end{align}
where
\ba
G_1(k_2,k_3,\theta_2,\theta_3 ) &=& -  \big[2 (k_2^2 +k_3^3) \, P_2(\cos \theta_3) + k_2 k_3 \cos{\theta_3}(1 + 3 \cos^2{\theta_3})\big] \\
G_2(k_2,k_3,\theta_2,\theta_3 ) &=&  \big[ k_2^2 \, P_2 (\cos {\theta_3}) 
+ 2 k_2 k_3 \, P_3(\cos{\theta_3}) +  k_3^2 \, P_4 ( \cos{\theta_3} ) \big] \cdot  P_6(\cos \theta_2)
\ea
and the $P_n$'s are the Legendre polynomials. All quantities that only depend on the magnitudes $k_1$, $k_2$, $k_3$ factor out of the $ \theta_2$ integral, because they cannot depend on the orientation of the triangle defined by the momenta.

We clearly see the two limiting cases now. For $\beta =9/2$, only the $G_1$ contribution survives, it has no $\theta_2$ dependence, and we recover the results of the isotropic solid inflation case. On the other hand, for $\beta = 8$, $G_1$ is gone and $G_2$, being proportional to $P_6(\cos \theta_2)$, averages to zero when we integrate over $\theta_2$.

\section{Anisotropic tensor spectrum?}
The existence of an anisotropic six-index invariant tensor suggests that anisotropies can also show up in the tensor modes' {\em two}-point function, because of the possible quadratic Lagrangian term
\be \label{Tgamma2} 
 T_{\rm aniso}^{ijklmn} \, \pa_i \gamma_{jk} \, \pa_l \gamma_{mn} \; .
\ee
However, it is easy to convince oneself that such a term cannot arise from expanding the lowest-derivative action we have been working with so far,
\be
S = \int d^4 x \sqrt{-g} \big[ \sfrac12 \mpl^2 R + F(B^{IJ}) \big] \; ,
\ee
simply because all possible anisotropies are in the structure of $F$, but its argument $B^{IJ} = g^{\mu\nu} \pa_\mu \phi^I \pa \phi^J$ does not involve derivatives of the metric. 

On the other hand, in the presence of higher derivative terms, one will generically get such a term. Consider for instance the invariant
\be \label{hd}
(g \cdots g )^{\mu_1 \nu _1 \cdots \mu_6 \nu_6 } \cdot \nabla_{\mu_1} \nabla_{\nu_1} \phi^{I_1} \cdots \nabla_{\mu_6} \nabla_{\nu_6} \phi^{I_6} \cdot T_{\rm aniso}^{I_1 \cdots I_6} \; ,
\ee 
where $(g \cdots g)$ stands schematically for any twelve-index tensor built out of the metric.  Setting the $\phi^I$'s to their background values $x^I$, and expanding in powers of the tensor modes $\gamma$, the covariant derivatives $\nabla \nabla \phi^I$ have the schematic form
\be
\nabla \nabla \phi^I \sim H + \pa \gamma \; ;
\ee
and so, upon taking all the contractions in \eqref{hd}, one does expect to find the term \eqref{Tgamma2} at quadratic order.
Similar considerations apply to higher-derivative terms that involve higher powers of curvature tensors, for instance a trilinear term schematically of the form
\be \label{hd2}
(R^{\mu\nu\rho\sigma}  \, \pa_\mu \phi^I \pa_\nu \phi^J \pa_\rho \phi^K \pa_\sigma \phi^L)^3 \; ,
\ee
with suitable contractions with our anisotropic invariant tensor $T_{\rm aniso}^{I_1 \cdots I_6}$. (We need at least three Riemann tensors, because our $T_{\rm aniso}$ is totally symmetric, while $R^{\mu\nu\rho\sigma}$ has antisymmetry properties as well.)

However, if we want the anisotropic quadratic terms that we get from these higher derivative corrections to compete with the purely isotropic ones we get from the Einstein-Hilbert action, we need to give the higher derivative corrections a large coefficient, of order $\mpl^2/H^4$ in the examples above. This makes the smallness of $F$,
\be
F \sim H^2 \mpl^2 \ll \mpl^4 \; ,
\ee
potentially unstable against quantum corrections. For instance, we expect graviton loops involving $\phi$'s on the external legs and the coupling \eqref{hd2} in the vertices, to  drastically correct $F(B^{IJ})$. A quick order-of-magnitude estimate of this two-loop diagram 
\be \nonumber
\includegraphics[width=5cm]{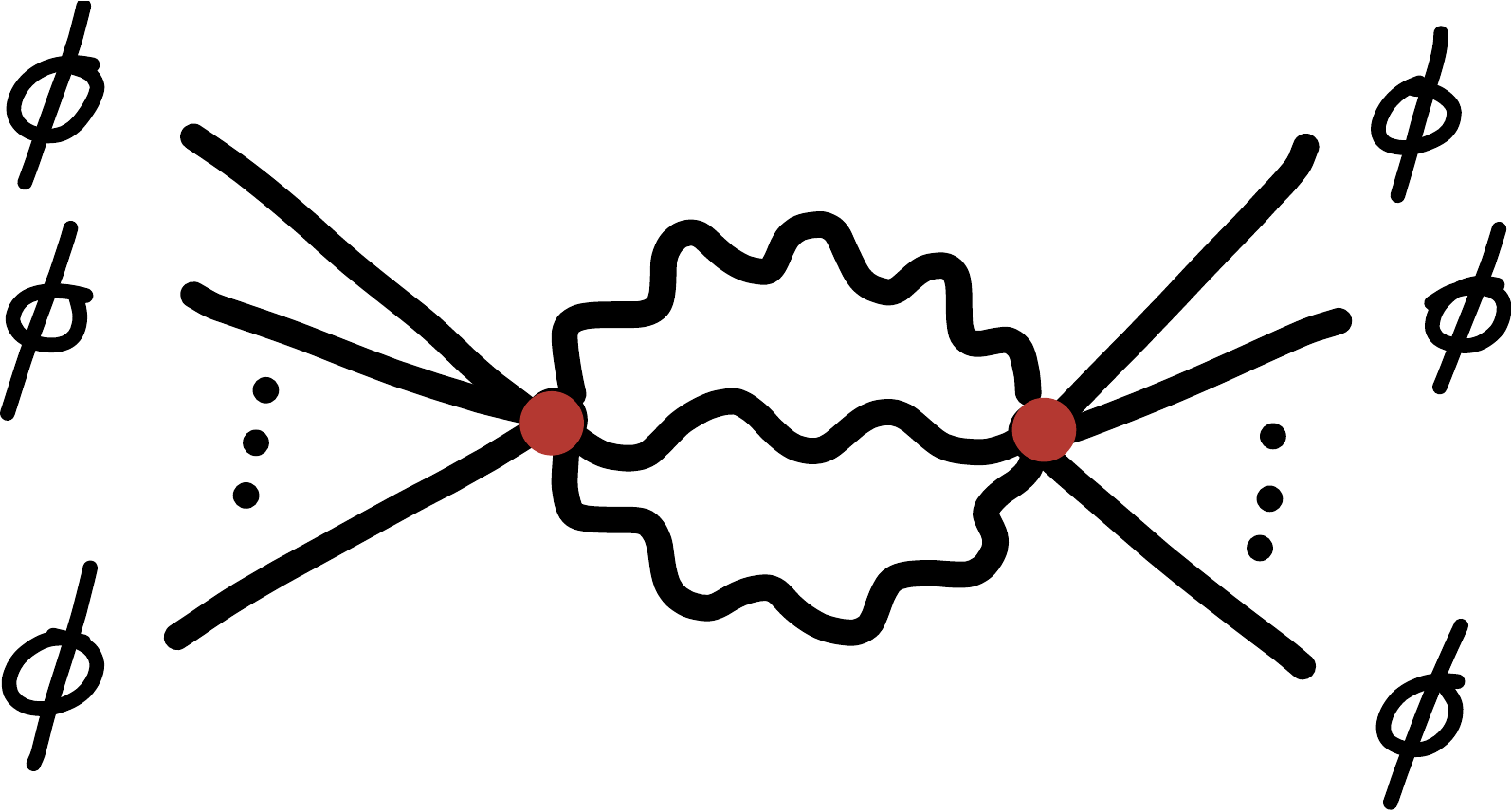} 
\ee
gives a correction to the effective action
\be
\Delta F \sim \frac{\mpl^5}{H} \epsilon^{21/2} \; ,
\ee
assuming the phonon speeds are relativistic, $c_L, c_T \sim 1$, and we cutoff the loop integrals at the solid's strong coupling scale, $\Lambda_{\rm strong} \sim \epsilon^{3/4} F^{1/4}$ \cite{Endlich:2012pz}. For $\Delta F$ to be at most of order $F$, we need a small enough $\epsilon$:
\be
\epsilon \lesssim \big( H/\mpl \big)^{2/7} \; .
\ee
This goes in the opposite direction to the bound on $\epsilon$ that guarantees that perturbations are weakly coupled at freeze out \cite{Endlich:2012pz},
\be
\epsilon \gg \big( H/\mpl \big)^{2/3} \; ,
\ee
but, for small $H/\mpl$, it is perfectly compatible with it.

We will not attempt a systematic analysis of higher derivative corrections here. However, the above estimates suggest that it may be consistent to expect higher derivative corrections  to be large enough to yield order-one anisotropies in the tensor spectrum, but  small enough to preserve the technical naturalness of our effective field theory. We will investigate the issue further in a forthcoming publication, where we will also give the detailed form of the resulting anisotropic tensor spectrum.

\section{Concluding remarks}
We have shown that, among the generalizations of solid inflation with discrete rotational symmetries, that with icosahedral symmetry is the only one that is naturally compatible with the observed isotropy of the background and of the scalar spectrum. 

The associated scalar three-point function is in general highly anisotropic, and this suppresses its overlap with all the standard templates used in CMB data analyses.  For a specific choice of the Lagrangian coefficients ($\beta = 8$, in our notation), it is {\em completely} anisotropic, in the sense that such an overlap vanishes exactly. This leaves open the possibility that a large non-gaussian signal is hiding in the data, waiting to be unveiled by a dedicated anisotropic analysis.

It is worth pointing out that our anisotropies are not of the same nature as those discussed in \cite{peloso, peloso2}: there, for any given realization one expects small anisotropies in the scalar spectrum; but, statistically speaking, these average to zero. On the other hand, in our case it is the statistical correlation functions themselves that are intrinsically anisotropic, potentially maximally so.

Similar considerations apply to the tensor spectrum as well: in the presence of sizable higher-derivative corrections, it can be highly anisotropic, which makes the standard detection strategies inefficient, and calls for a dedicated analysis.

For the scalar modes, we see no reason for the $\beta = 8$ case to be preferred over others; for instance we see no symmetry protecting it against quantum corrections. However, it is a simple, consistent limit of our theory, and we find it interesting that such a completely anisotropic limit exists at all. Is it an accidental feature of our truncation of the theory at the cubic/three-point function level, or does it survive at higher orders as well?

\section*{Acknowledgements}
We would like to thank Paolo Creminelli, Lam Hui, Eiichiro Komatsu, Federico Piazza, and Kendrick Smith for useful discussions and comments. We are especially indebted to Solomon Endlich and Riccardo Penco for collaboration in the early stages of this project. Our work is partially supported by the DOE under contract no.~DE-FG02-11ER41743 and by the Kwanjeong Educational Foundation.

\section*{Appendix}

\appendix

\section{Background stress-energy tensor}

To relate the background stress-energy tensor to $F$ and its derivatives, it is useful to organize $F$'s dependence on $B^{IJ}$ in terms of the variables
\be \label{X and b}
X \equiv [B] \; , \qquad b^{IJ} \equiv \frac{B^{IJ}}{[B]}   \qquad \qquad (B^{IJ} = g^{\mu\nu} \, \pa_\mu \phi^I \pa_\nu \phi^J) \; .
\ee
$X$ depends on the overall normalization of $B^{IJ}$ whereas $b^{IJ}$ does not. As a result, the approximate internal scale invariance \eqref{scale} translates into a weak $X$-dependence of $F$.

Taking the variation with respect to the metric for the solid action
\be \label{solid action}
S_{\rm solid} = \int d^4 x \, \sqrt{-g} \, F(X, b^{IJ}) \; , 
\ee
we find the stress-energy tensor
\be \la{emtensor}
T_{\mu\nu} = g_{\mu\nu}F - 2 F_X \pa_\mu \phi^I \pa_\nu \phi^I - \frac{2}{X} F_{IJ} \big( \pa_\mu \phi^I \pa_\nu \phi^J - b^{IJ} \pa_\mu \phi^K \pa_\nu \phi^K \big) \; ,
\ee
where the subscript $X$ and $IJ$ stand for partial derivatives w.r.t.~$X$ and $b^{IJ}$. 

When we evaluate $T_{\mu\nu}$ on the background configuration, we can use the fact that $F_{IJ}$ must be icosahedral invariant. As we saw, for a two-index tensor this implies that it is proportional to $\delta_{IJ}$. The terms in parentheses in \eqref{emtensor} thus cancel against each other, and we are left with the same background stress-energy tensor as in $SO(3)$-invariant solid inflation. In particular:
\begin{align}
& \rho = -F,\quad p=F-\tfrac{2}{a^2} F_X
\\
& \epsilon = \fr{F_X X}{F} \; . 
\end{align}

\section{Phonon propagation speeds}

To find the phonon propagation speeds, we should expand the solid action \eqref{solid action} to quadratic order in the phonon field $\vec \pi$. Let's use the same $X$ and $b^{ij}$ variables of last section; the expansion of the action then reads
\be \label{L phonon}
{\cal L} = F_X \, \delta X + F_{ij} \, \delta b^{ij} + \sfrac 12 F_{XX} (\delta X)^2 + F_{X,ij} \, \delta X \, \delta b^{ij} + \sfrac 12 F_{ij,kl} \,  \delta b^{ij} \delta b^{kl} + \dots
\ee
When we specialize all the derivatives of $F$ to the background, by icosahedral symmetry they must take the form
\be
F_{ij}, F_{X,ij} \propto \delta_{ij} \; , \qquad F_{ij,kl} = f_1 \, \delta_{ij} \delta_{kl} + f_2  \, \big( \delta_{ik} \delta_{jl} + \delta_{il} \delta_{jk} \big) \; ,
\ee
with generic, time-dependent coefficients. This kills some of the terms in \eqref{L phonon} because, 
by definition (eq.~\eqref{X and b}), the fluctuation of $b^{ij}$ is traceless. We are left with
\be
{\cal L} \simeq F_X \, \delta X + \sfrac 12 F_{XX} (\delta X)^2 + f_2 \,  (\delta b^{ij})^2
\ee

We thus need $\delta X$ up to quadratic order in the phonon field, and $\delta b^{ij}$ up to linear order. These are
\be
\delta X = \pi^{ii} \; , \qquad \delta b^{ij} \simeq \sfrac13 \big(\pi^{ij}  - \sfrac13 \pi^{kk} \delta^{ij} \big)  \; ,
\ee
where $\pi^{ij}$ is the fluctuation of $B^{ij}$,
\be
B^{ij} = \delta^{ij} + \pi^{ij} \; , \qquad \pi^{ij} = \pa^i \pi^j + \pa^j \pi^i + \pa_\mu \pi^i \pa^\mu \pi^j \; .
\ee
At quadratic order in $\vec \pi$ we get
\be
{\cal L}_2 =  -F_X \big[ \, \dot{\vec{\pi}}^2 - c_T^2 (\pa_i \pi_j )^2 -(c_L^2 - c_T^2)(\nabla \cdot \vec{\pi})^2 \, \big]
\ee
with
\be
c_L^2 = 1 + {2}\fr{F_{XX}}{F_X} + \fr{8}{27}\fr{f_2}{F_X}, \quad c_T^2 \equiv 1 + \fr{2}{9} \fr{f_2}{F_X}
\ee

This is identical to the $SO(3)$-invariant solid inflation's result, upon identifying
\be
f_2 \big|_{\rm here} \quad \leftrightarrow \quad (F_Y+F_Z)\big|_{\rm there} \; .
\ee
As a consequence, icosahedral inflation still obeys the relation \eqref{cL cT}, and, more in general, is indistinguishable from solid inflation at the quadratic level.


\bibliographystyle{utphys_jonghee}
\bibliography{icosahedral}{}

\providecommand{\href}[2]{#2}\begingroup\raggedright\begin{thebibliography}{10}

\bibitem{Endlich:2012pz}
S.~Endlich, A.~Nicolis, and J.~Wang, ``{Solid Inflation},''
  \href{http://dx.doi.org/10.1088/1475-7516/2013/10/011}{{\em JCAP} {\bfseries
  1310} (2013) 011},
\href{http://arxiv.org/abs/1210.0569}{{ arXiv:1210.0569 [hep-th]}}.

\bibitem{steinhardt}
D.~Levine and P.~J. Steinhardt, ``Quasicrystals: A new class of ordered
  structures,'' \href{http://dx.doi.org/10.1103/PhysRevLett.53.2477}{{\em Phys.
  Rev. Lett.} {\bfseries 53} (Dec, 1984) 2477--2480}.

\bibitem{DGNR}
S.~Dubovsky, T.~Gregoire, A.~Nicolis, and R.~Rattazzi, ``{Null energy condition
  and superluminal propagation},''
  \href{http://dx.doi.org/10.1088/1126-6708/2006/03/025}{{\em JHEP} {\bfseries
  03} (2006) 025},
\href{http://arxiv.org/abs/hep-th/0512260}{{ arXiv:hep-th/0512260 [hep-th]}}.

\bibitem{gruzinov}
A.~Gruzinov, ``{Elastic inflation},''
  \href{http://dx.doi.org/10.1103/PhysRevD.70.063518}{{\em Phys. Rev.}
  {\bfseries D70} (2004) 063518},
\href{http://arxiv.org/abs/astro-ph/0404548}{{ arXiv:astro-ph/0404548
  [astro-ph]}}.

\bibitem{sigurdson}
M.~Sitwell and K.~Sigurdson, ``{Quantization of Perturbations in an Inflating
  Elastic Solid},'' \href{http://dx.doi.org/10.1103/PhysRevD.89.123509}{{\em
  Phys. Rev.} {\bfseries D89} (2014) 123509},
\href{http://arxiv.org/abs/1306.5762}{{ arXiv:1306.5762 [astro-ph.CO]}}.

\bibitem{Babich:2004gb}
D.~Babich, P.~Creminelli, and M.~Zaldarriaga, ``{The Shape of
  non-Gaussianities},''
  \href{http://dx.doi.org/10.1088/1475-7516/2004/08/009}{{\em JCAP} {\bfseries
  0408} (2004) 009},
\href{http://arxiv.org/abs/astro-ph/0405356}{{ arXiv:astro-ph/0405356
  [astro-ph]}}.

\bibitem{litvin}
D.~B. Litvin, ``{The icosahedral point groups},'' {\em Acta Crystallographica
  Section A} {\bfseries 47} (Mar, 1991) 70--73.

\bibitem{Cheung:2007st}
C.~Cheung, P.~Creminelli, A.~L. Fitzpatrick, J.~Kaplan, and L.~Senatore, ``{The
  Effective Field Theory of Inflation},''
  \href{http://dx.doi.org/10.1088/1126-6708/2008/03/014}{{\em JHEP} {\bfseries
  03} (2008) 014},
\href{http://arxiv.org/abs/0709.0293}{{ arXiv:0709.0293 [hep-th]}}.

\bibitem{Endlich:2013jia}
S.~Endlich, B.~Horn, A.~Nicolis, and J.~Wang, ``{Squeezed limit of the solid
  inflation three-point function},''
  \href{http://dx.doi.org/10.1103/PhysRevD.90.063506}{{\em Phys.Rev.}
  {\bfseries D90} (2014) 063506},
\href{http://arxiv.org/abs/1307.8114}{{ arXiv:1307.8114 [hep-th]}}.

\bibitem{peloso}
N.~Bartolo, S.~Matarrese, M.~Peloso, and A.~Ricciardone, ``{Anisotropy in solid
  inflation},'' \href{http://dx.doi.org/10.1088/1475-7516/2013/08/022}{{\em
  JCAP} {\bfseries 1308} (2013) 022},
\href{http://arxiv.org/abs/1306.4160}{{ arXiv:1306.4160 [astro-ph.CO]}}.

\bibitem{peloso2}
N.~Bartolo, M.~Peloso, A.~Ricciardone, and C.~Unal, ``{The expected anisotropy
  in solid inflation},''
  \href{http://dx.doi.org/10.1088/1475-7516/2014/11/009}{{\em JCAP} {\bfseries
  1411} (2014) 009},
\href{http://arxiv.org/abs/1407.8053}{{ arXiv:1407.8053 [astro-ph.CO]}}.

\end{thebibliography}\endgroup

\end{document}